\documentclass{iopjournal}
\usepackage{lineno,hyperref}
\usepackage[cp1251]{inputenc}
\usepackage[numbers]{natbib}
\usepackage{xcolor}
\usepackage{latexsym,amsmath,amssymb}
\usepackage{amsbsy,graphicx}
\usepackage{epstopdf}
\usepackage{epsfig}

\begin{document}

\articletype{paper} 

\title{Frenkel line of Yukawa fluids within \\ the self-consistent relaxation theory\textsuperscript{*}}

\author{Ilnaz I. Fairushin$^1$\orcid{0000-0002-6721-7488}
and Anatolii V. Mokshin$^{1,*}$\orcid{0000-0003-2919-864X}
}

\affil{$^1$Department of Computational Physics, Institute of Physics, Kazan Federal University, 420008 Kazan, Russia}

\affil{$^*$Author to whom any correspondence should be addressed.}

\email{anatolii.mokshin@mail.ru}


\renewcommand{\thefootnote}{\fnsymbol{footnote}}
\footnotetext[1]{\it In memory of Prof. Kostya Trachenko.}

\begin{abstract}
For a simple model fluid, the Yukawa fluid, the condition for dynamic crossover, known as the Frenkel line, is defined. The condition is related to the fact that the roton minimum in the dispersion relation of longitudinal acoustic-like excitations exists only when the collective vibrational dynamics of particles dominates in the liquid. Based on the self-consistent relaxation theory for the Yukawa fluid, thermodynamic states are determined in which the roton minimum disappears. The obtained values of the state parameters for the Frenkel line are consistent with the results of studies in which the position of this line on the phase diagram of the Yukawa fluid was determined using  molecular dynamics simulations. It is shown that the Frenkel line in this system can be determined directly from the structural characteristic -- the static structure factor. A physical interpretation of the roton minimum frequency for the simple liquids near the Frenkel line is proposed.
\end{abstract}

\newpage


A supercritical fluid is a condensed substance characterized by the absence of a pronounced transition between liquid and gaseous phases \cite{KTrachenko_book, Cowan}. This state is achieved when both temperature and pressure exceed their critical values, which can be defined only for systems that exhibit a liquid-vapor phase transition. Supercritical fluids are of great interest because they have wide applications in various fields \cite{McHardy}. In recent decades, transition/crossover lines have been identified in the thermodynamic region of a supercritical fluid, separating regions with distinct thermodynamic and purely dynamic regimes. Examples include the Widom line \cite{Stanley}, the Fisher-Widom line \cite{Fisher_JPC}, and the Frenkel line \cite{KTrachenko_book, KTrachenko_PRE_2012, Baggioli, Ferracina_2026, Feng_2026, JPCM_FL_Review_2018, JCP_2026_CO2, PRR_2026_Electronic fluids, JPC_B_2025, PRB_2017_Trachenko, ACS omega_2023, PRE_2014_Trachenko, JPCL_2022_water, SR_2019_LJ, JPC_B_2018, PRE_2015_Trachenko, PR_2021_Trachenko}.

The Frenkel line is a dynamic crossover line on the phase diagram of a substance in a supercritical state. It separates two regions with qualitatively distinct particle behavior: gaslike dynamics, where particles move almost freely, and solidlike dynamics, where their motion has an activation character governed by collective interactions.
According to Ref. \cite{KTrachenko_PRE_2012}, the Frenkel line is defined as a set of points on the phase diagram at which the Frenkel relaxation time $\tau_R$ is approximately equal to the minimum (Debye) period of atomic oscillations in the liquid $\tau_D$ \cite{Frenkel_book}:
\begin{equation}\label{gen_cond_FL}
\tau_R \approx \tau_D.
\end{equation}
Note that $\tau_R$ decreases with increasing temperature $T$, while $\tau_D$ depends weakly on temperature and is mainly governed by interparticle interactions \cite{KTrachenko_PRE_2012}.
\begin{figure}[h!]
\centering
\includegraphics[width=7.0cm]{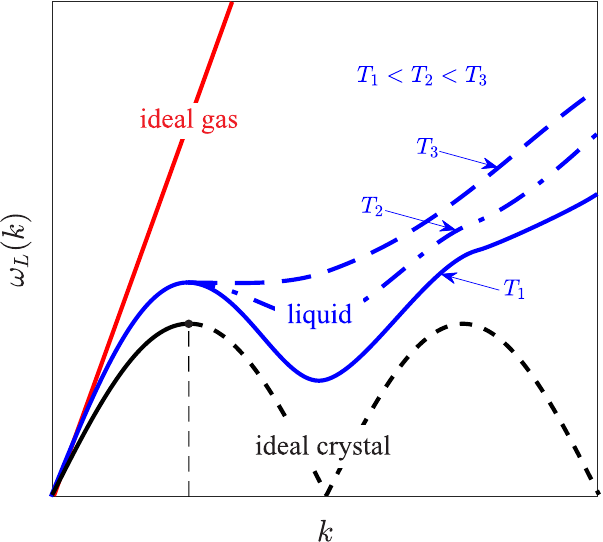}
\caption{Schematic representation of the dispersion relation of longitudinal collective excitations for various states of matter: crystal, liquid, and gas. The vertical dashed line indicates the boundary of the first Brillouin zone.}
\label{Scheme}
\end{figure}

Of particular interest are systems of particles with purely repulsive interactions. Such systems lack a sharp liquid-vapor phase transition, and under appropriate thermodynamic conditions they directly correspond to the definition of supercritical fluid \cite{Balucani, Hansen/McDonald_book_2006, Khrapak_Review_2024}. Importantly, the absence of a liquid-vapor phase transition in systems with purely repulsive interparticle interactions dramatically changes the topology of their phase diagrams, including the location of dynamic crossovers. There is no Widom line, whereas the Frenkel line and Fisher-Widom line are not tied to the critical point, which is absent in such the systems. An important example of systems with purely repulsive interactions between particles is the Yukawa fluid. Firstly, the Yukawa many-particle systems have been actively studied for many years in the context of describing the physical properties of strongly coupled and dusty plasmas \cite{Baggioli, Barrat, Ichimaru, Hamaguchi, Fortov_reviews_1, Fairushin, Fluids, Mithen_PRE_2011, Arkhipov, TkachenkoPRE2020, BraultPRE2025, MurilloPRR2022}. Secondly,  using the method of moments \cite{Arkhipov, TkachenkoPRE2020} and the self-consistent relaxation theory \cite{AVM_PRE_2001, MokshinTMF, Mokshin, PRB_2020}, the key problem of statistical mechanics was recently solved for the case of the Yukawa fluid: a theoretical description of dynamic properties in terms of structural characteristics was developed \cite{MFT_PRE, FM_PRE, FM_PRE_2025}. The main obtained result here was an analytical expression for the spectra of the dynamic structure factor $S(k, \omega)$, which reproduces  correctly all the features of collective particle dynamics over a wide range of thermodynamic states. In addition, an analytical expression was derived for the longitudinal collective excitation dispersion relation $\omega_L(k)$. This is significant point because, in the context of the Frenkel line, the dispersion relation $\omega_L(k)$ in liquids exhibits a characteristic feature: the presence or absence of a roton minimum in the boundary region between the second and third Brillouin pseudozones, depending on the particle dynamics regime \cite{KTrachenko_PRE_2012, Wang_2019, Kalman_CPP, Trigger_2026} (see Fig.~\ref{Scheme}). When crossing the Frenkel line from a state with predominantly vibrational particle dynamics to a state with predominantly diffusive particle dynamics, the roton minimum vanishes. Thus, examining the dependence of the dispersion relation $\omega_L(k)$ on the state parameters allows one to determine the thermodynamic parameters corresponding to the Frenkel line of the Yukawa fluid. A study of the roton minimum in dispersion relation of longitudinal collective excitations $\omega_L(k)$ has a rich history \cite{Godfrin_PRB_2021, Wang_2019, Kalman_CPP, Trigger_2026}. As known for superfluid helium, collective excitations with wave numbers corresponding to the roton minimum are well-defined eigenmodes and they decay very weakly \cite{Godfrin_PRB_2021, Trigger_2026}. In the case of simple classical liquids, to which the Yukawa fluid also belongs, collective excitations with wave numbers near the roton minimum are not eigenmodes and they decay rapidly \cite{MFT_PRE}. Here, the roton minimum characterizes the smearing of the boundary of the second pseudo-Brillouin zone and is not associated with collective excitations in a strict sense \cite{Wang_2019}. The roton minimum in a classical liquid can be interpreted as an averaged frequency dispersion of local microcrystals (cages) formed by strongly correlated particles \cite{Kalman_CPP}.

\begin{figure}[h!]
\centering
\includegraphics[width=7.7cm]{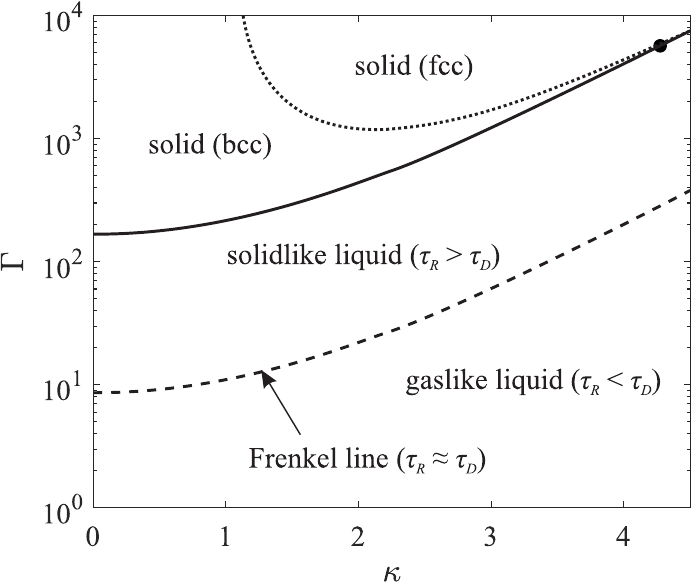}
\caption{Phase diagram of the Yukawa system \cite{Hamaguchi}. The solid line corresponds to the melting line, and the dashed line indicates the Frenkel line according to Ref. \cite{Baggioli}. The circle marks the triple point. In Ref. \cite{Baggioli}, an empirical correspondence between the melting and Frenkel lines was proposed: $\Gamma_{\rm F}(\kappa)/\Gamma_{\rm m}(\kappa)=0.05$.}
\label{PD_FL1}
\end{figure}

A thermodynamic state of the Yukawa system is setted by the coupling parameter
\begin{equation}
\Gamma = \frac{Q^2}{4\pi\varepsilon_0ak_BT}
\label{Gamma}
\end{equation}
and the Debye screening parameter
\begin{equation}
\kappa=\frac{a}{\lambda_s}.
\end{equation}
Here, $a = (3/4\pi \rho)^{1/3}$ is the radius of the Wigner-Seitz cell, $\rho$ is the number density of the particles, $\varepsilon_0$ is the electric constant, $k_B$ is the Boltzmann constant, $T$ is the absolute temperature. Further, $\lambda_s$ is the Debye screening length, which determines the softness of the Yukawa interaction potential
\begin{equation}\label{Yukawa_pot}
u(r) = \frac{Q^2}{4\pi\varepsilon_0 r}\exp\left(-\frac{r}{\lambda_s}\right).
\end{equation}
In the limit $\lambda_s\rightarrow\infty$, the potential (\ref{Yukawa_pot}) reduces to the bare Coulomb potential, and in the opposite limit $\lambda_s\rightarrow0$, it approximates the hard-sphere
interaction potential \cite{Fortov_reviews_1}. In the case of the three-dimensional Yukawa system, the melting line is given to a good accuracy by the following expression (Fig.~\ref{PD_FL1}) \cite{Vaulina_1}:
\begin{equation}
\Gamma_{\rm m}(\kappa) = \frac{172\exp{(\alpha\kappa)}}{0.5(\alpha\kappa)^2+\alpha\kappa+1},
\end{equation}
where $\Gamma_{\rm m}$ is the value of the coupling parameter at which the liquid-solid phase transition occurs, $\alpha=(4\pi/3)^{1/3}$.
As can be seen from the phase diagram of the Yukawa system (Fig.~\ref{PD_FL1}), in the liquid region, the transition from a regime with predominantly solidlike particle dynamics ($\tau_R > \tau_D$) to a regime with gaslike dynamics ($\tau_R < \tau_D$) occurs with a decrease in the coupling parameter $\Gamma$ or with an increase in the screening parameter $\kappa$.

\begin{figure}[h!]
\centering
\includegraphics[width=0.9\linewidth]{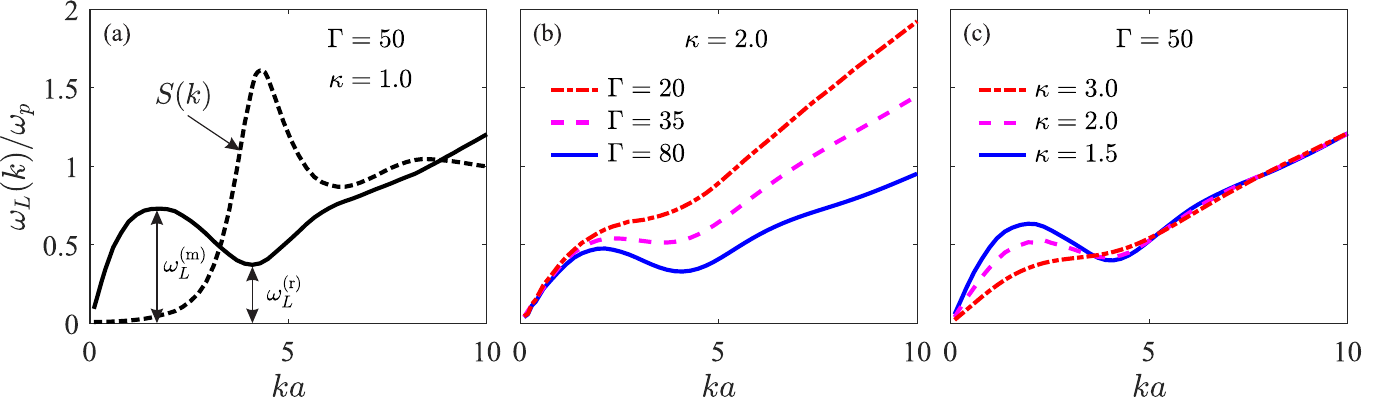}
\caption{Dispersion curves of longitudinal collective excitations $\omega_L(k)$ normalized by the plasma frequency $\omega_p$ for the Yukawa fluid at different ($\Gamma,\kappa$)-states, calculated using the self-consistent relaxation theory via Eq.~(\ref{w_L}).}
\label{Disp_wL}
\end{figure}

The dispersion relation $\omega_L(k)$ is determined by the main peak position of the longitudinal current spectral density $C_L(k, \omega)$, which is related to the dynamic structure factor $S(k, \omega)$ as
\begin{equation}
C_{L}(k,\omega )=\frac{3\Gamma\omega^2}{(\omega_pka)^2}S(k,\omega ).
\label{eq: CL}
\end{equation}
According to the basic definition, the dynamic structure factor $S(k,\omega)$ is the Fourier transform (in time) of the density fluctuations time correlation function $F(k,t)$ \cite{Balucani, Hansen/McDonald_book_2006}:
\begin{equation}\label{Skw_from_Fkt}
S(k,\omega) = \frac{S(k)}{2\pi}\int_{-\infty}^\infty F(k,t)\exp(\textbf{i}\omega t)dt\,,
\end{equation}
where $S(k)$ is the static structure factor, $t$ is the time. In turn, the time correlation function $F(k,t)$ can be expanded in the Taylor series as
\begin{equation}
F(k,t) = 1 - \langle\omega^{(2)}(k)\rangle\frac{t^2}{2!}+\langle\omega^{(4)}(k)\rangle\frac{t^4}{4!}+\dots +(-\textbf{i})^l\langle\omega^{(l)}(k)\rangle\frac{t^l}{l!}+\dots \,,
\label{Fkt_series}
\end{equation}
where $\langle\omega^{(l)}(k)\rangle$ is the normalized frequency moment of $S(k, \omega)$ of the $l$th order ($l = 2, 4 , 6, ...$):
\begin{equation}
\langle\omega^{(l)}(k)\rangle = (-\textbf{i})^l\frac{d^l}{dt^l}F(k,t)\bigg|_{t=0}= \frac{1}{S(k)} \int_{-\infty}^\infty \omega^l S(k,\omega)d\omega\,.
\label{nmoments_basic}
\end{equation}
From Eq. (\ref{Fkt_series}), one obtains the following expression for the Laplace transform of the function $F(k,t)$:
\begin{equation}
\widetilde{F}(k, s) = \frac{1}{s} - \frac{\langle\omega^{(2)}(k)\rangle}{s^3} + \frac{\langle\omega^{(4)}(k)\rangle}{s^5} +\dots + (-\textbf{i})^l\frac{\langle\omega^{(l)}(k)\rangle}{s^{l+1}} +\dots\,,
\label{Fks_series}
\end{equation}
which can be rewritten as the continued fraction:
\begin{equation}
\widetilde{F}(k,s) = \cfrac{1}{s+\cfrac{\Delta_1(k)}{s+\cfrac{\Delta_2(k)}{s+\cfrac{\Delta_3(k)}{s+\ddots}}}} \,.
\label{Fks_cfrac}
\end{equation}
Here, $\Delta_n(k)$ are the frequency relaxation parameters \cite{ AVM_PRE_2001, MokshinTMF, Mokshin, PRB_2020}. The frequency relaxation parameters and the frequency moments are related as follows \cite{MFT_PRE, FM_PRE}
\begin{flalign}
\Delta _{1}(k) &=\frac{\langle \omega ^{(2)}(k)\rangle }{\langle \omega
^{(0)}(k)\rangle }, \notag \\
\Delta _{2}(k) &=\frac{\langle \omega ^{(4)}(k)\rangle }{\langle \omega
^{(2)}(k)\rangle }-\frac{\langle \omega ^{(2)}(k)\rangle }{\langle \omega
^{(0)}(k)\rangle },  \notag \\
\Delta _{3}(k) &=\frac{\left[ \langle \omega ^{(6)}(k)\rangle \langle
\omega ^{(2)}(k)\rangle -\left( \langle \omega ^{(4)}(k)\rangle \right) ^{2}%
\right] \langle \omega ^{(0)}(k)\rangle }{\langle \omega ^{(4)}(k)\rangle
\langle \omega ^{(2)}(k)\rangle \langle \omega ^{(0)}(k)\rangle -\left(
\langle \omega ^{(2)}(k)\rangle \right) ^{3}},  \notag \\
\,\,\, \dots, \notag \\
\Delta _{n}(k) &= \mathcal{F}\left[ \langle \omega ^{(0)}(k)\rangle, \langle
\omega ^{(2)}(k)\rangle, \dots , \langle \omega ^{(2n)}(k)\rangle \right],
\end{flalign}
where $\mathcal{F}[\ldots]$ means an algebraic expression. Further, the $n$-th order frequency relaxation parameter $\Delta _{n}(k)$ is defined by the corresponding distribution function of $n$ particles of the system and characterizes the oscillatory process for different groups of $n$ neighboring particles \cite{AVM_PRE_2001, MokshinTMF, Mokshin, PRB_2020, MFT_PRE, FM_PRE}. On the other hand, the quantity $\tau_n(k) = 1/\sqrt{\Delta_n(k)}$ determines the time scale of the corresponding relaxation process. The first four quantities $\tau_1(k)$, $\tau_2(k)$, $\tau_3(k)$ and $\tau_4(k)$ of the set $\{ \tau_n(k) \}$ correlate with the time scales of the known hydrodynamic variables.
The self-consistent relaxation theory invokes a reduction to a limited set of relevant dynamical variables based on the assumption that the time scales of the higher-order dynamical variables are equalized \cite{AVM_PRE_2001, MokshinTMF, Mokshin, PRB_2020, MFT_PRE, FM_PRE}, i.e.,
\begin{equation}
\label{Deltas_equal}
1/\sqrt{\Delta_4(k)} = 1/\sqrt{\Delta_5(k)} = 1/\sqrt{\Delta_6(k)} = \ldots .
\end{equation}
As a result, fraction (\ref{Fks_cfrac}) will be presented just in terms of the first four frequency relaxation parameters, $\Delta_1(k)$, $\Delta_2(k)$, $\Delta_3(k)$ and $\Delta_4(k)$. The following analytical expressions are known for $\Delta_1(k)$ and $\Delta_2(k)$ \cite{Mithen_PRE_2011, MFT_PRE}:
\begin{equation}
\Delta_1(k) = \frac{\omega_p^2(ka)^2}{3\Gamma S(k)} \,,
\label{delta_1_micro}
\end{equation}
\begin{eqnarray} \label{eq: Delta_2}
\Delta _{2}(k) &=& \Delta _{1}(k)\left[ 3S(k)-1\right]  \nonumber\\
               &+& \omega _{p}^{2} \int_{0}^{\infty }\frac{\exp (-\kappa x)}{3x}\bigl[ \left( 2(\kappa
                    x)^{2}+6\kappa x+6\right) j_{2}(kax) \nonumber\\
               & &\hskip 3cm + (\kappa x)^{2}(1-j_{0}(kax))\bigr]g(x)dx.
\end{eqnarray}
Here, $\omega_p=\sqrt{Q^2\rho/(\varepsilon_0m)}$ is the plasma
frequency, $x=r/a$ is the dimensionless distance, $j_{n}(x)$ is the spherical Bessel functions of the first kind, and $g(x)$ is the radial distribution function. Further, the third- and fourth-order frequency relaxation parameters can be expressed through the parameter $\Delta_2(k)$ as follows (Ref. \cite{MFT_PRE})
\begin{subequations}
\label{eq: freq_parameters_approx}
\begin{equation}
\Delta_3(k) = \frac{3}{2}\Delta_2(k) + \omega_0^2,
\label{approx1}
\end{equation}
\begin{equation}
\Delta_4(k) = \frac{4}{3}\Delta_3(k) = 2\Delta_2(k) + \frac{4}{3}\omega_0^2,
\label{approx2}
\end{equation}
\end{subequations}
where
\begin{equation}
\omega_0^2 = \frac{2\,\omega_p^2}{\sqrt{\Gamma\kappa}}.
\nonumber
\end{equation}
Taking into account Eqs. (\ref{Fks_cfrac}), (\ref{Deltas_equal}) and the relation
\begin{equation}
S(k,\omega) = \frac{S(k)}{\pi}\operatorname{Re}\left[\widetilde{F}(k,s=\textbf{i}\omega)\right],
\end{equation}
one obtains the analytical expression for the dynamic structure factor
\begin{equation}\label{Skw_MSC}
S(k,\omega)=\frac{S(k)}{\pi }\frac{2\Delta _{2}(k)\sqrt{A_{3}(k)}}{\omega
^{6}+A_{1}(k)\omega ^{4}+A_{2}(k)\omega ^{2}+A_{3}(k)},
\end{equation}
where
\begin{flalign}
A_{1}(k) &=3\omega _{0}^{2}-\frac{\Delta _{2}(k)}{2}-2\Delta _{1}(k),
\notag \\
A_{2}(k) &=\left[ \Delta _{1}(k)-2\Delta _{2}(k)\right] ^{2}-6\Delta
_{1}(k)\omega _{0}^{2},  \notag \\
A_{3}(k) &=\frac{3}{2}\Delta _{1}^{2}(k)\left( 3\Delta _{2}(k)+2\omega
_{0}^{2}\right) .  \notag
\end{flalign}
Eqs. (\ref{eq: CL}) and (\ref{Skw_MSC}) yield the dispersion equation
\begin{equation}\label{w_L_eq}
2\omega^6_L(k) + A_1(k)\omega^4_L(k) + A_2(k) = 0
\end{equation}
with the solutions
\begin{equation} \label{w_L}
\omega^2_L(k) = C_+(k) + C_-(k) - \frac{A_1(k)}{6},
\end{equation}
where
\begin{equation}
C^3_\pm(k) = \frac{1}{4}\left(A_2(k) - \frac{A_1^3(k)}{54} \pm \sqrt{ A_2^2(k) - \frac{A_2(k)A_1^3(k)}{27}}\right).
\end{equation}
Let $\omega_L^{(\rm m)}$ and $\omega_L^{(\rm r)}$ be denoted as the heights of the local maximum and the roton minimum of the $\omega_L(k)$ dispersion curves, respectively (Fig.~\ref{Disp_wL}a). Then, the dispersion relations $\omega_L(k)$ for the ($\Gamma,\kappa$)-states corresponding to the Frenkel line satisfy the following condition:
\begin{equation}\label{FL_disp_cond}
\omega_L^{(\rm m)}=\omega_L^{(\rm r)}.
\end{equation}
Note that since the dispersion relation $\omega_L(k)$ can be determined directly from the static structure factor $S(k)$ via Eq.~(\ref{w_L}), the parameters $\Gamma_{\rm F}$ and $\kappa_{\rm F}$ corresponding to the Frenkel line can be obtained from structural data alone using condition (\ref{FL_disp_cond}). In the present work, the static structure factor $S(k)$ is obtained via the Fourier transform of the radial distribution function $g(r)$, which determined from  molecular dynamics simulations\footnote{The equilibrium molecular dynamics simulations of the Yukawa fluid were carried out using the computational package LAMMPS [S. Plimpton, J. Comput. Phys. 117,1 (1995)]. The simulation cell contained 64 000 particles interacting with the Yukawa potential, and periodic boundary conditions were applied to the cell in all directions. The evolution of the system corresponding to the NVT ensemble was monitored. The particle motion equations were integrated in accordance with the Verlet algorithm with a time integration step $\tau = 0.01/\omega_p$.}:
\begin{equation}\label{Sk_from_gr}
S(k)=1+\frac{4\pi\rho}{k}\int_0^\infty r\left(g(r)-1\right)\sin(kr)dr.
\end{equation}
Using Eq. (\ref{w_L}), the dispersion relation $\omega_L(k)$ is obtained for various ($\Gamma, \kappa$)-states at fixed coupling parameter $\Gamma$. The $\omega_L(k, \kappa)$-surface is then approximated by splines, and an extremum analysis is performed to determine the screening parameter $\kappa$ at which the roton minimum disappears, i.e., when condition (\ref{FL_disp_cond}) is satisfied. As found, the error of this numerical procedure does not exceed 5\%.

Fig.~\ref{Disp_wL} shows dispersion curves for longitudinal collective excitations in the Yukawa fluid, obtained from Eq.~(\ref{w_L_eq}) for thermodynamic states with different particle dynamics regimes. One can see that the location of the roton minimum correlates with the main maximum of the static structure factor $S(k)$ (Fig.~\ref{Disp_wL}a), i.e., it lies at the boundary of the second Brillouin pseudozone. As expected, at a fixed screening parameter $\kappa$, the roton minimum becomes less pronounced with decreasing coupling parameter $\Gamma$ (Fig.~\ref{Disp_wL}b), which corresponds to an increase in temperature (see Eq.~(\ref{Gamma})). This occurs because the interaction energy of a particle with its neighbors becomes small compared to its thermal energy, leading to a reduction of the relaxation time $\tau_R$ -- that is, particles jump more frequently between temporary equilibrium positions \cite{Frenkel_book}. At a fixed coupling parameter $\Gamma$, the dispersion curves $\omega_L(k)$ exhibit a weaker roton minimum as the screening parameter $\kappa$ increases (Fig.~\ref{Disp_wL}c). This effect arises from the fact that with increasing $\kappa$ the Yukawa interaction potential (\ref{Yukawa_pot}) becomes shorter-ranged, and as a consequence the particles' motion becomes less constrained, while the Debye time $\tau_D$ grows.

\begin{figure}[h!]
\centering
\includegraphics[width=7.7cm]{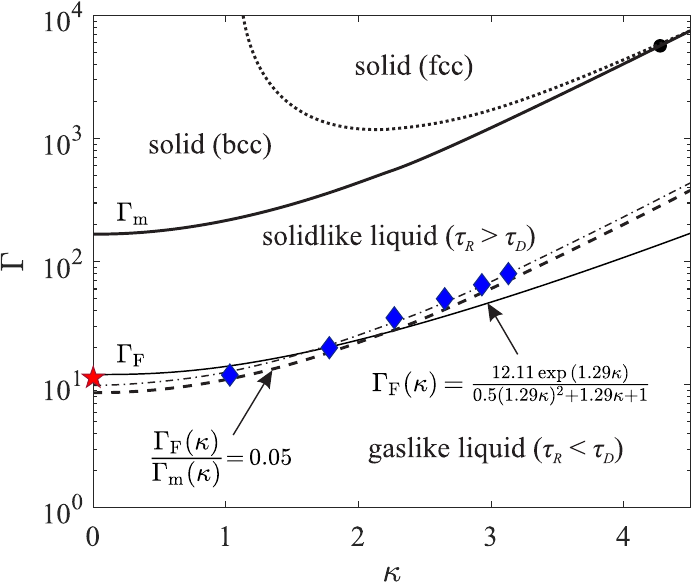}
\caption{Phase diagram of the Yukawa system \cite{Hamaguchi}. The solid line is the melting line; the dashed and thin solid lines are the Frenkel lines for the Yukawa fluid from Refs. \cite{Baggioli} and \cite{Feng_2026}, respectively. The circle marks the triple point. The remaining symbols indicate ($\Gamma, \kappa$)-states for which the condition (\ref{FL_disp_cond}) is satisfied in the dispersion curves of longitudinal collective excitations according to the self-consistent relaxation theory. The thin dash-dotted line is the best fit to these data and corresponds to the expression $\Gamma_{\rm F}(\kappa)/\Gamma_{\rm m}(\kappa)=0.057$.}
\label{PD_FL2}
\end{figure}

The ($\Gamma,\kappa$)-states corresponding to the Frenkel line according to condition (\ref{FL_disp_cond}) are shown in Fig.~\ref{PD_FL2}. This figure also displays the state with $\Gamma=11.42$ and $\kappa=0$, for which, according to the self-consistent relaxation theory, the roton minimum vanishes on the longitudinal collective excitation dispersion curves of the Coulomb system \cite{FM_PRE}. The line on the phase diagram (Fig.~\ref{PD_FL2}) that best fits the results of the present work is given by $\Gamma_{\rm F}/\Gamma_{\rm m}=0.057$. At the same time, the Frenkel line obtained from molecular dynamics (MD) simulations by analyzing various thermodynamic and transport characteristics corresponds to the ratio $\Gamma_{\rm F}/\Gamma_{\rm m}=0.05$ \cite{Baggioli}. As seen, condition (\ref{FL_disp_cond}) allows one to determine the Frenkel line of simple liquids from an experimentally measurable quantity -- the dispersion relation of longitudinal collective excitations. For the Yukawa fluid, this approach requires only the static structure factor $S(k)$. In this context, the recent work \cite{Feng_2026} is of particular interest. Using the isomorph theory, the authors of this work define the Frenkel line for two- and three-dimensional Yukawa fluids and conclude that, for the three-dimensional case, the Frenkel line corresponds to thermodynamic states where the first peak height of the radial distribution function is $g(x_{\rm max})=1.14$. As found by the authors, this criterion deviates significantly from the condition $\Gamma_{\rm F}/\Gamma_{\rm m}=0.05$ (see Fig.~\ref{PD_FL2}). Since the present work mainly deals with Yukawa fluid states at $\kappa<3$, it is difficult to draw a definite conclusion as to how the approach of Ref. \cite{Feng_2026} relates to ours. Note that although correlation relations (\ref{approx1}) and (\ref{approx2}) were obtained for coupling and screening parameters in the ranges $\Gamma \in [20, 100]$ and $\kappa \in [1, 2]$ (see Ref. \cite{MFT_PRE}), the results of the present work demonstrate their validity also beyond these ranges.

Acoustic-like longitudinal excitations with wavelengths below twice the interatomic distance (the first Brillouin zone boundary, Fig.~\ref{Scheme}) cannot propagate in a crystal. The maximum of dispersion $\omega_L(k)$ thus gives the limiting frequency of longitudinal phonons in the crystal. By analogy, for liquid states, where the vibrational dynamics of particles dominates, the height of the local maximum $\omega_L^{(\rm m)}$ can be regarded as the maximum frequency of longitudinal collective excitations. As the Frenkel line is approached from the solid-like side, transverse collective excitations vanish, leaving only longitudinal ones \cite{KTrachenko_PRE_2012}. Therefore, in these thermodynamic states $\omega_L^{(\rm m)}$ can be identified with the Debye frequency of particle vibrations \cite{KTrachenko_PRE_2012, Khrapak_Review_2024}. Consequently, near the Frenkel line the following relation holds:
\begin{equation}
\tau_D\approx 1/\omega_L^{(\rm m)}.
\end{equation}
Next, since conditions (\ref{gen_cond_FL}) and (\ref{FL_disp_cond}) hold on the Frenkel line, we obtain the following relation for the Frenkel relaxation time $\tau_R$:
\begin{equation}\label{tau_R_omega_L}
\tau_R \approx 1/\omega_L^{(\rm r)}.
\end{equation}
The validity of the last expression is also supported by the following two arguments. First, the roton minimum always lies to the right of the maximum in $\omega_L(k)$, i.e., at larger wave-numbers, and its position corresponds to spatial scales below twice the interparticle distance. It is, therefore, governed by the intrinsic vibrational dynamics of particles in the cage of nearest neighbors, rather than by collective excitations, whose wavelengths are bounded below by twice the interparticle distance. Second, the dependence of the frequencies $\omega_L^{(\rm m)}$ and $\omega_L^{(\rm r)}$ of the longitudinal dispersion curves of the Yukawa fluid on the states parameters $\Gamma$ and $\kappa$ reveals the following. For a fixed $\kappa$, the roton minimum height $\omega_L^{(\rm r)}$ varies more strongly with the coupling parameter $\Gamma$ than the maximum height $\omega_L^{(\rm m)}$ does (Fig.~\ref{Disp_wL}b). Hence, $\omega_L^{(\rm r)}$ exhibits a pronounced temperature sensitivity, suggesting a connection with the inverse Frenkel relaxation time $1/\tau_R$. In contrast, for a fixed $\Gamma$, the maximum height $\omega_L^{(\rm m)}$ depends more strongly on the screening parameter $\kappa$ than the roton minimum height $\omega_L^{(\rm r)}$ does (Fig.~\ref{Disp_wL}c). Hence, $\omega_L^{(\rm m)}$ is sensitive to the softness of the Yukawa potential (\ref{Yukawa_pot}), which predominantly governs collective excitations on the scale of the inverse Debye period $1/\tau_D$. The arguments leading to Eq.~(\ref{tau_R_omega_L}) pertain to the dispersion relation $\omega_L(k)$ in general and can therefore be applied to various many-particle systems. However, verifying the validity of Eq.~(\ref{tau_R_omega_L}) requires, in addition to the known dispersion relation $\omega_L(k)$, the determination of the quantity $\tau_R$. Such an analysis is planned for various systems in future studies.

Thus, this paper demonstrates that knowledge of the dispersion relation for longitudinal collective excitations in simple liquids allows one to determine the particle dynamics regime and the location of the Frenkel line on the phase diagram. We illustrate this approach using the Yukawa fluid within the self-consistent relaxation theory of collective dynamics. The sets of thermodynamic states corresponding to the Frenkel line obtained by the proposed method are consistent with MD simulations results. We also show that, near the Frenkel line, the roton minimum frequency in the dispersion relation of longitudinal collective excitations is directly related to the Frenkel relaxation time.

\section*{Acknowledgments}
This work/publication was funded by a grant from the Academy of Sciences of the Republic of Tatarstan provided to higher education institutions, scientific and other organizations to support human resource development plans in terms of encouraging their research and academic staff to defend doctoral dissertations and conduct research activities (Agreement No. 12/2025-PD-KFU dated December 22, 2025).


\begin{thebibliography}{99}
\bibitem{KTrachenko_book} K. Trachenko, {\it Theory of Liquids: From Excitations to Thermodynamics} (Cambridge University Press, Cambridge, 2023).

\bibitem{Cowan} J. Cowan, J. Cann, {\sl Nature} {\bf 333}, 259 (1988).

\bibitem{McHardy} J. McHardy, S. P. Sawan, {\it Supercritical Fluid Cleaning: Fundamentals, Technology and Applications} (Noyes Publications, New Jersey, 1998).

\bibitem{Stanley} L. Xu, P. Kumar, S. V. Buldyrev, S.-H. Chen, P. H. Poole, E. Sciortino, H. E. Stanley, {\sl Proc. Natl. Acad. Sci. USA} {\bf 102}, 16558 (2005).

\bibitem{Fisher_JPC} M. E. Fisher, B. Wiodm, {\sl J. Chem. Phys.} {\bf 50}, 3756 (1969).

\bibitem{KTrachenko_PRE_2012} V. V. Brazhkin, Yu. D. Fomin, A. G. Lyapin, V. N. Ryzhov, K. Trachenko, {\sl Phys. Rev. E} {\bf 85}, 031203 (2012).

\bibitem{Baggioli} D. Huang, M. Baggioli, S. Lu, Z. Ma, Y. Feng, {\sl Phys. Rev. Research} {\bf 5}, 013149 (2023).

\bibitem{Ferracina_2026} F. Ferracina, A. K. A Lu, X. N. Du, N. Chen, M. I. Ojovan, H. Suito, D. V. Louzguine-Luzgin, {\sl Journal of Physics: Condensed Matter} {\bf 38}, 205702 (2026).

\bibitem{Feng_2026} A. Xu, N. Yu, D. Huang, Y. Feng, {\sl Phys. Rev. Research} {\bf 8}, 023016 (2026).

\bibitem{JPCM_FL_Review_2018} Y. D. Fomin, V. N. Ryzhov, E. N. Tsiok, J. E. Proctor, C. Prescher, V. B. Prakapenka, K. Trachenko, V. V. Brazhkin,  {\sl Journal of Physics: Condensed Matter} {\bf 30}, 134003 (2018).


\bibitem{JCP_2026_CO2}	Y. Li, J. Zhang, X. Liu, Y. Zeng, Z. Jin, H. Zhang, {\sl Journal of Chemical Physics} {\bf 165}, 064507 (2026).

\bibitem{PRR_2026_Electronic fluids}	J. Fournier, P. O. Downey, O. Gingras, C. D. H\'{e}bert, M. Charlebois, A. M. Tremblay, {\sl Phys. Rev. Research} {\bf 8}, 023278 (2026).

\bibitem{JPC_B_2025} C. G. Pruteanu, A. D. Daramola, M. Kirsz, C. E. Robertson, L. J. Jones, T.  Wang, J. S. Loveday, G. J. Ackland, O. L. G. Alderman, J. E. Proctor, {\sl Journal of physical chemistry. B} {\bf 129}, 3420 (2025).

\bibitem{PRB_2017_Trachenko} C. Prescher, Y. D. Fomin, V. B. Prakapenka, J. Stefanski, K. Trachenko, V. V. Brazhkin, {\sl Phys. Rev. B} {\bf 95}, 134114 (2017).

\bibitem{ACS omega_2023} C. G. Pruteanu, M. N. Bannerman, M. Kirsz, L. Lue, G. J. Ackland, {\sl ACS omega} {\bf 8}, 12144 (2023).

\bibitem{PRE_2014_Trachenko} K. Trachenko, V. V. Brazhkin, and D. Bolmatov, {\sl Phys. Rev. E} {\bf 89}, 032126 (2014).

\bibitem{JPCL_2022_water} I. Skarmoutsos, J. Samios, E. Guardia, {\sl Journal of Physical Chemistry Letters} {\bf 13}, 7636 (2022).

\bibitem{SR_2019_LJ} K. Ghosh, C. V. Krishnamurthy, {\sl Scientific Reports} {\bf 9}, 14872 (2019).

\bibitem{JPC_B_2018} J. E. Proctor, M. Bailey, I. Morrison, M. A. Hakeem, I. F. Crowe, {\sl Journal of physical chemistry. B} {\bf 122}, 10172 (2018).

\bibitem{PRE_2015_Trachenko} C. Yang, V. V. Brazhkin, M. T. Dove, K. Trachenko, {\sl Phys. Rev. E} {\bf 91}, 012112 (2015).

\bibitem{PR_2021_Trachenko} C. Cockrell, V. V. Brazhkin, K. Trachenko, {Phys. Rep.} {\bf 941}, 1 (2021).

\bibitem{Frenkel_book} J. Frenkel, {\it Kinetic Theory of Liquids} (Oxford University Press, New York, 1946).

\bibitem{Balucani} U. Balucani, M. Zoppi, {\it Dynamics of the Liquid State} (Oxford University Press, New York, 2002).

\bibitem{Hansen/McDonald_book_2006} J.-P. Hansen, I. R. McDonald, \textit{Theory of Simple Liquids} (Academic Press, London, 2006).

\bibitem{Khrapak_Review_2024} S. A. Khrapak, {\sl Phys. Rep.} {\bf 1050}, 1 (2024).

\bibitem{Barrat} J. L. Barrat, J. P. Hansen, H. Totsuji,  {\sl Journal of Physics C: Solid State Physics} {\bf 21}, 4511 (1988).

\bibitem{Ichimaru} S. Ichimaru,  {\sl Rev. Mod. Phys.} {\bf 65}, 255 (1993).

\bibitem{Hamaguchi} S. Hamaguchi, R. T. Farouki, D. H. E. Dubin, {\sl Phys. Rev. E} {\bf 56}, 4671 (1997).

\bibitem{Fortov_reviews_1} V. E. Fortov, A. V. Ivlev, S. A. Khrapak, A. G.
Khrapak, G. E. Morphill, {\sl Phys. Rep.} {\bf 421}, 1 (2005).

\bibitem{Fairushin} I. I. Fairushin, S. A. Khrapak, A. V. Mokshin, {\sl Results in Physics} {\bf 19}, 103359 (2020).

\bibitem{Fluids} I. I. Fairushin, A. V. Mokshin, {\sl Fluids} {\bf 8}, 72 (2023).

\bibitem{Mithen_PRE_2011} J. P. Mithen, J. Daligault, J. B. Crowley, G.
Gregori, {\sl Phys. Rev. E} {\bf 84}, 046401 (2011).

\bibitem{Arkhipov} Yu. V. Arkhipov, A. Askaruly, A. E. Davletov, D. Yu. Dubovtsev, Z. Donko, P. Hartmann, I. Korolov, L. Conde, I. M. Tkachenko, {\sl Phys. Rev. Lett.} {\bf 119}, 045001 (2017).

\bibitem{TkachenkoPRE2020} Y. V. Arkhipov, A. Ashikbayeva, A. Askaruly, A. E. Davletov, D. Y. Dubovtsev, K. S. Santybayev, S. A. Syzganbayeva, L. Conde, I. M. Tkachenko, {\sl Phys. Rev. E} {\bf 102}, 053215 (2020).

\bibitem{BraultPRE2025} S. S. Mishra, S. Bhattacharjee, P. Brault, {\sl Phys. Rev. E} {\bf 111}, 015208 (2025).

\bibitem{MurilloPRR2022} D. Huang, S. Lu, M. S. Murillo, Y. Feng, {\sl Phys. Rev. Research} {\bf 4}, 033064 (2022).

\bibitem{AVM_PRE_2001} R. M. Yulmetyev, A. V. Mokshin, P. H\"{a}nggi, V.Yu. Shurygin, {\sl Phys. Rev. E} {\bf 64}, 057101 (2001).

\bibitem{MokshinTMF}  A. V. Mokshin,  {\sl Theor. Math. Phys.} {\bf 183}, 449 (2015).

\bibitem{Mokshin} A. V. Mokshin, B. N. Galimzyanov, {\sl J. Phys.: Condens. Matter} {\bf 30}, 085102 (2018).

\bibitem{PRB_2020} R. M. Khusnutdinoff, C. Cockrell, O. A. Dicks, A. C. S. Jensen, M. D. Le, L. Wang, M. T. Dove, A. V. Mokshin, V. V. Brazhkin, K. Trachenko, {\sl Phys. Rev. B} {\bf 101}, 214312 (2020).

\bibitem{MFT_PRE} A. V. Mokshin, I. I. Fairushin, I. M. Tkachenko {\sl Phys. Rev. E} {\bf 105}, 025204 (2022).

\bibitem{FM_PRE} I. I. Fairushin, A. V. Mokshin, {\sl Phys. Rev. E} {\bf 108}, 015206 (2023).

\bibitem{FM_PRE_2025} I. I. Fairushin, A. V. Mokshin, {\sl Phys. Rev. E} {\bf 112}, 015210 (2025).

\bibitem{Godfrin_PRB_2021} H. Godfrin, K. Beauvois, A. Sultan, E. Krotscheck, J. Dawidowski, B. F\aa k, J. Ollivier, {\sl Phys. Rev. B} {\bf 103}, 104516 (2021).

\bibitem{Wang_2019} L. Wang, C. Yang, M. T. Dove, A. V. Mokshin, V. V. Brazhkin, K. Trachenko, {\sl Scientific Reports} {\bf 9}, 755 (2019).

\bibitem{Kalman_CPP} G. J. Kalman, S. Kyrkos, K. I. Golden, P. Hartmann, Z. Donko, {\sl Contrib. Plasma Phys.} {\bf 52}, 219 (2012).

\bibitem{Trigger_2026} S. A. Trigger, {\sl Journal of Low Temperature Physics} {\bf 222}, 61 (2026).

\bibitem{Vaulina_1} O. S. Vaulina, S. A. Khrapak, {\sl J. Exp. Theor. Phys.} {\bf 90}, 287 (2000).

\end{thebibliography}
\end{document}